\documentclass[doublecol,final,a4]{epl2} 

\usepackage{amssymb}
\usepackage{graphicx}
\usepackage{amsmath}
\usepackage{feynmp}
\usepackage{color}
\usepackage{hyperref}
\usepackage{floatrow}
\usepackage{widetable}
\newfloatcommand{capbtabbox}{table}[][\FBwidth]

\newcommand{\rr}{\mathbf{r}}

\newcommand{\kk}{\mathbf{k}}

\newcommand{\GG}{\mathcal{G}}

\newcommand{\rhom}{R_{\mathrm{TF,0}}}

\newcommand{\eps}{\varepsilon}
\newcommand{\neql}{n_\mathrm{eq}}
\newcommand{\Beq}{\mathcal{B}_\mathrm{eq}}
\newcommand{\VM}[1]{VM${}_\text{#1}$}

\title{Universal behavior of repulsive two-dimensional fermions in the vicinity of the quantum freezing point}
\shorttitle{Strongly correlated quasi-two-dimensional dipolar fermions}

\author{Mehrtash Babadi\inst{1} \and Brian Skinner\inst{2} \and Michael M. Fogler\inst{3} \and Eugene Demler\inst{1}}
\shortauthor{M. Babadi \etal}

\institute{                    
  \inst{1} Physics Department, Harvard University, Cambridge, Massachusetts 02138, USA\\
  \inst{2} School of Physics and Astronomy, University of Minnesota, Minneapolis, Minnesota 55455, USA\\
  \inst{3} Department of Physics, University of California San Diego, La Jolla, California 92093, USA
}

\abstract{
We show by a meta-analysis of the available Quantum Monte-Carlo (QMC) results that two-dimensional fermions with repulsive interactions exhibit universal behavior in the strongly-correlated regime, and that their freezing transition can be described using a quantum generalization of the classical Hansen-Verlet freezing criterion. We calculate the liquid-state energy and the freezing point of the 2D dipolar Fermi gas (2DDFG) using a variational method by taking ground state wave functions of 2D electron gas (2DEG) as trial states. A comparison with the recent fixed-node diffusion Monte-Carlo analysis of the 2DDFG shows that our simple variational technique captures more than $95\%$ of the correlation energy, and predicts the freezing transition within the uncertainty bounds of QMC. Finally, we utilize the ground state wave functions of 2DDFG as trial states and provide a variational account of the effects of finite 2D confinement width. Our results indicate significant beyond mean-field effects. We calculate the frequency of collective monopole oscillations of the quasi-2D dipolar gas as an experimental demonstration of correlation effects.}

\begin{document}

\maketitle

\begin{table*}
  \begin{widetable}{\columnwidth}{lcllcccc} \hline
  $V(r)$ & Statistics & Critical Coupling & $k_\mathrm{max}/k_F$ & $\gamma_m$ &  $S_f$ & $M_\mathrm{roton}/m$ & Reference(s)\\ \hline
  $e^2/r$ & Fermion & $r_s = 28(3)$ & $1.86$ & 0.25(-) & $1.53(4)$ & 0.042(4) & \cite{Tanatar1989,Attaccalite2002,Drummond2009,Chui1991}\\
  $D^2/r^3$ & Fermion & $g_d = 25(3)$ & $1.89$ & - & $1.55(3)$ & 0.041(5) & \cite{Matveeva2012}\\
  $(\sigma/r)^\infty$ & Fermion & $\tilde{\sigma} = 1.62(4)$ & $1.93$ & - & $1.54(3)$ & 0.043(2) & \cite{Drummond2011}\\  
  $-e^2\,\ln r$ & Boson & $r_s = 12(1)$ & - & 0.24(1) & - & - & \cite{Magro1994}\\
  $D^2/r^3$ & Boson & $g_d = 60(3)$ & 1.87 & 0.230(6) & 1.70(4) & 0.061(5) & \cite{Astrakharchik2007}\\
  $(\sigma/r)^\infty$ & Boson & $\tilde{\sigma} = 2.00(5)$ & 1.93 & 0.279 & 1.54(-) & 0.065(-) & \cite{Xing1990}\\ 
  $\epsilon K_0(\sigma/r)$ & Boson & cf. Ref.~\cite{Magro1993} & - & 0.235(15) & - & - & \cite{Magro1993}\\ 
\hline
\end{widetable}
\caption{Characteristics of single-component 2D quantum liquids at the freezing point. $\gamma_m$ is the Lindemann ratio at melting, $k_\mathrm{max}$ is the location of the main peak of the static structure factor $S(k)$, $k_F \equiv \sqrt{4\pi n}$ is the Fermi wavevector, and $M_\mathrm{roton}$ is the roton mass calculated from the Bijl-Feynman formula. The dashed entries could not be determined from the available data. The error bounds reflect both the statistical error of fitting/interpolating as well as the uncertainty in the value of the critical coupling. The dimensionless coupling constants are defined as: $r_s \equiv m e^2/(\hbar^2 \sqrt{\pi n})$ (the Wigner-Seitz radius), $g_d \equiv k_F m D^2/\hbar^2$ for dipolar interactions ($m$ and $D$ denote the mass and dipole moment of a single particle), and $\tilde{\sigma} \equiv k_F \sigma$ for hard-core gases ($\sigma$ is the diameter of the hard-core disk).}
\end{table*}

An intriguing behavior of fermions with strong repulsive interactions is the spontaneous breaking of the translation symmetry in the ground-state and the formation of the so-called Wigner crystal 
(WC) phase. While originally proposed for the electron gas~\cite{Wigner1934}, a large body of evidence from Quantum Monte-Carlo (QMC) simulations and first-principle considerations have shown that the WC transition is indeed a universal aspect of repulsively interacting particles, independent of their quantum statistics, number of spatial dimensions, interaction law or spin degeneracy. Some of the extensively studied models that exhibit the WC transition are the electron gas in 2D~\cite{Tanatar1989,Attaccalite2002,Drummond2009} and 3D~\cite{Ceperley1980}, 2D Coulomb bosons~\cite{Magro1994}, 2D Yukawa bosons~\cite{Magro1993}, 2D dipolar bosons\cite{Astrakharchik2007} and fermions~\cite{Matveeva2012}, and 2D hard-core bosons~\cite{Xing1990} and fermions~\cite{Drummond2011}.

The conventional explanation of WC transition at zero temperature is based on the competition between quantum fluctuations (kinetic energy) and the inter-particle repulsion, favoring delocalized and localized states, respectively. The symmetry broken state is energetically favorable when the ratio of the interaction over kinetic energy becomes sufficiently large. The ordered state is a triangular crystal in 2D which has the largest packing ratio. While general arguments from the Landau-Ginzburg theory suggest that the WC transition is a direct first-order transition, for interaction laws falling slower than $1/r^3$~\cite{Spivak2004}, the first-order transition may be replaced by a series of second-order transitions through intermediate ``microemulsion'' phases such as stripes and bubbles~\cite{Jamei2005}. Such transitions, however, generally take place only over a very narrow window of densities and remain yet to be observed in QMC simulations due to finite size limitations~\cite{Clark2009,Matveeva2012}.

In this Letter, we investigate the features of the strongly-correlated liquid phase of fermions in the vicinity of the WC transition and show that models with significantly different interaction laws exhibit universal features. For concreteness, we restrict our analysis to single-component fermions in 2D. Throughout this letter, we use the terminology ``universal'' to refer to properties that depend very weakly on the microscopic interaction laws. As a first step, we present a meta-analysis of the available QMC studies of different models and point out their universal and non-universal features. As a useful application and a token of evidence for the universality (in the above sense) of strongly-correlated liquid-state wave functions, we estimate the energy and the WC transition point of 2D dipolar fermionic gas (2DDFG) using 2D electron gas (2DEG) wave functions as variational trial states. We show that the obtained results are in a remarkable agreement with the available QMC results~\cite{Matveeva2012}. Finally, we make a connection to experiments with ultracold dipolar gases and investigate the effects of strong correlations in quasi-2D dipolar fermions in optical traps. Our results suggest an unexpectedly weak dependence of the total energy and the collective modes on the transverse confinement width.

\section{Universality in the ``roton regime''}
\begin{figure}
\centering
\includegraphics[scale=0.62]{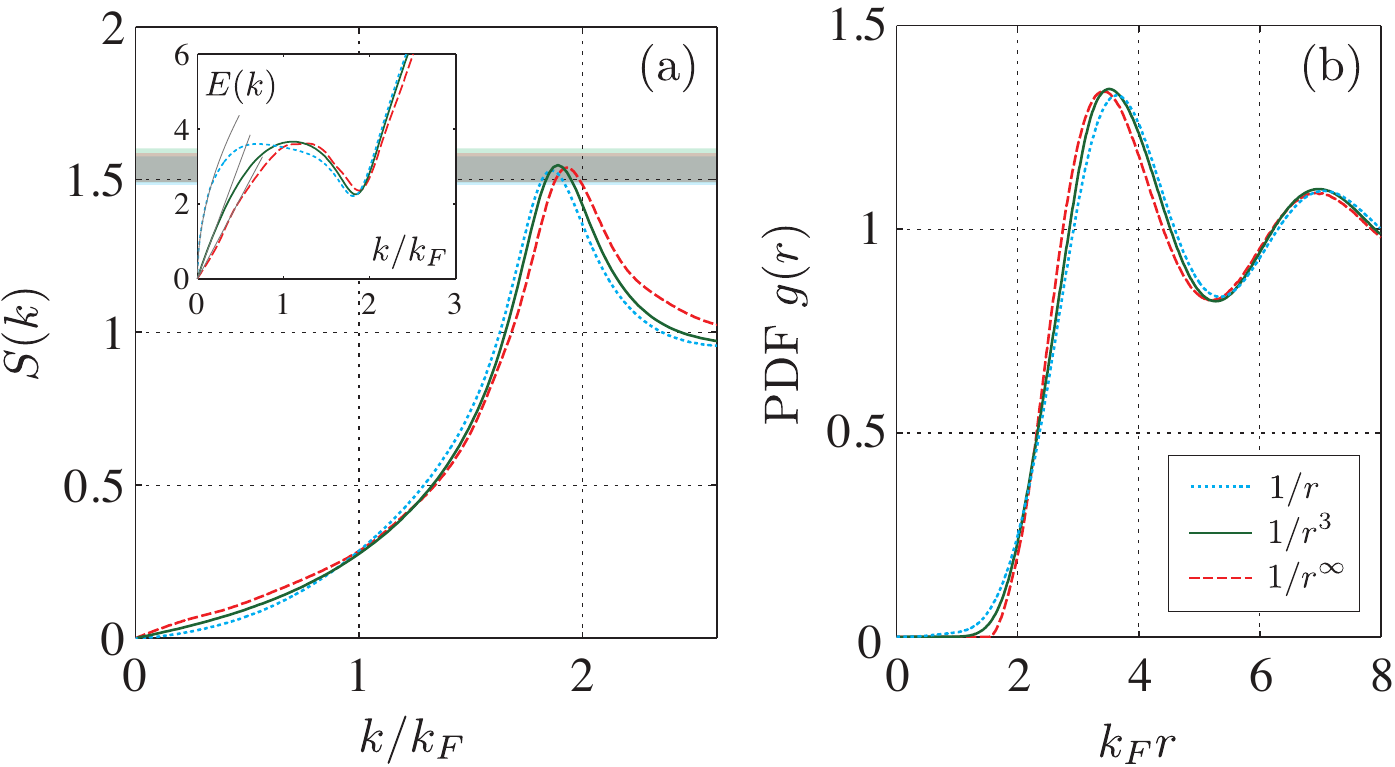}
\caption{(a) The liquid state static structure factor $S(k)$ for 2D electron gas, dipolar fermions and hard-core fermions at the critical point. The horizontal bar indicates the uncertainty in $S(k_\mathrm{max})$. The inset plot shows the upper-bound dispersion of the collective modes $E(k)$ from Bijl-Feynman formula (in the units of $\varepsilon_F = k_F^2/2m$). The roton softening at $k = k_\mathrm{max} \approx 2k_F$ and the different dispersion of long wavelength modes are visible. The thin gray lines in the inset plot are guide to the eye. (b) the pair-distribution function $g(r)$ at the critical point.}
\end{figure}

Classical liquid-solid phase transitions are known to follow universal patterns, such as the Lindemann criterion of melting and the Hansen-Verlet (HV) criterion of freezing~\cite{HansenMcDonald}. The former states that the solid melts once the Lindemann ratio $\gamma_L \equiv \langle u^2 \rangle^{1/2}/a_0$ exceeds a universal constant, $\gamma_m$ (here, $u$ is the particle displacement in the crystal lattice and $a_0$ is the lattice constant). The HV criterion states that the liquid freezes once the main peak of the static structure factor, $S(k_\mathrm{max})$, exceeds a universal constant $S_f$.
Despite the lack of a rigorous first-principle explanation, simulations and experiments have shown that both criteria are universally applicable to simple liquids with a $5\%$ to $10\%$ variation in the constants~\cite{HansenMcDonald}.

By analyzing the published QMC results for a variety of quantum models, we find that both criteria apply to quantum melting and freezing transitions as well, albeit at different values of $\gamma_m$ and $S_f$ compared to the classical case (see Table 1). The quantum generalization of the Lindemann criterion had been indicated before~\cite{Runge1988} and is often used heuristically. The most significant finding here is the universality of the HV constant for 2DEG, dipolar and hard-core fermions despite their fundamentally different interaction laws. This surprising result implies that although 2DEG and dipolar/hard-core fermions crystallize in the opposite low- and high-density regimes, they still comply with the same freezing criterion.

The universality of the HV constant suggests that the physics of 2D Fermi liquids in the regime $k \sim k_\mathrm{max} \approx 2k_F$ may be in fact universal near the freezing point. To shed light into this matter, we examine the liquid state static structure factor $S(k)$ and the associated pair distribution function (PDF) $g(r) = n^{-2}\langle n(\rr) n(0)\rangle \equiv 1 + \int\mathrm{d}^2\kk\,[S(k)-1]\,e^{i\kk\cdot\rr}/(4\pi^2 n)$ for 2DEG, dipolar and hard-core fermions at the critical point (Fig.~1). The inset plot of Fig.~1a shows an upper bound to the density-wave dispersions from the Bijl-Feynman single-mode approximation, $E(k) = k^2/[2m S(k)]$ (exact in the $k \ll k_F$ limit). Save for differences in the short- and long- wavelength regimes, the models manifestly exhibit a universal behavior in the intermediate ``roton regime'' $k \sim 2 k_F$: the same ``roton gap'' $E(k^*)$ and the same ``roton mass'' $M_\mathrm{roton} \equiv \hbar^2/E''(k^*)$ ($k^*$ is the location of the minimum of $E(k)$). Provided that the Bijl-Feynman expression remains reliable in the roton regime, the universality of the HV constant is equivalent to the universality of the roton gap at the transition.

The model-dependent features show up in the short wavelength $k \gg k_F$ and long wavelength $k \ll k_F$ regimes. The former is seen as different degrees of flatness of the plots of $g(r)$ in the $k_F r \ll 1$ regime (Fig.~1b). The ``pair amplitude'' $\sqrt{g(r)}$ obeys the two-body Schr\"odinger's equation in this limit~\cite{Kimball} and directly reflects the hardness of the repulsive core of the potential. The long wavelength behavior of the models is most easily seen in the density wave dispersion plots: while $E(k) \sim k$ (zero-sound/phonon) for the dipolar and hard-core fermions, $E(k) \sim k^{1/2}$ (plasmon) for 2DEG. While the non-universality of the long wavelength physics may play a significant role in the narrow critical regime (e.g. the stability of mesoscopic phases), it is immaterial as far as the thermodynamical stability of the pure phases is concerned.

The universal features of critical liquids in the roton regime are expected to persist in an extended neighborhood of the critical point, as long as the crystal correlations are strong. This statement will be examined and verified below, but for the moment, it can be motivated by observing that in the strongly-correlated regime, the short-range part of the interaction law ($k_F r \lesssim 1$) is effectively masked by the localized exchange-correlation hole~\cite{Wang1991}, i.e. the combined effects of Pauli exclusion and short-range inter-particle repulsion. Moreover, the momentum dependence of the power-law potentials is smooth and does not vary appreciably in the roton regime while the long-range tails have little contribution due to screening. As a consequence, the dispersion of the density waves in the roton regime is expected to have a weak dependence on the shape of the interaction potential.

A significant fraction of the total energy results from the roton regime in the strongly-correlated regime due to the reduced roton gap. Therefore, a practical application of the roton universality is that the ground-state wave functions of one model will be well-suited as variational trial states for another model. Such variational approaches has been used earlier by two of the authors~\cite{Fogler2005,Skinner2010}. The quality of the variational results, in particular in the strongly-correlated regime, constitutes a stringent test for the similarity between the wave functions of different systems. Here, we will estimate the ground state energy and the freezing point of 2DDFG by taking the ground state wave functions of 2DEG as trial states. The details of this procedure is described in the next section.

\section{The variational mapping method} 
Let $|\Psi_\xi\rangle^A$, $\eps^A_{\mathrm{K}}[n;\xi^A]$ and $g^A(r;\xi^A)$ be the normalized ground state wave function, kinetic energy per particle, and the PDF of a reference system ``A'' at a fixed density $n$. Here, $\xi^A$ is a dimensionless coupling constant of ``A'' (e.g. in case of 2DEG, $\xi$ can be taken as the Wigner-Seitz radius $r_s$). We use $\{|\Psi_\xi\rangle^A\}$ as a family of variational wave functions for the target system ``B''. Let $\eps^B_{\mathrm{K}}[n;\xi_\mathrm{var}]$ and $ \eps^B_{\mathrm{int}}[n;\xi_\mathrm{var}]$ be the kinetic and interaction energy densities of ``B'' obtained by using $|\Psi_{\xi_\mathrm{var}}\rangle^A$ as a trial wave function. Since the kinetic energy operator $\hat{\mathcal{K}} \equiv -\sum_i \hbar^2 \nabla_i^2/(2m)$ is identical for both systems, $\eps^B_\mathrm{K}[n;\xi_\mathrm{var}] \equiv {}^A\langle \Psi_{\xi_\mathrm{var}}|\hat{\mathcal{K}}|\Psi_{\xi_\mathrm{var}}\rangle^A = \eps^A_\mathrm{K}[n;\xi_\mathrm{var}]$. The interaction energy per particle of ``B'' in the same trial state can be calculated using the PDF of ``A'':
\begin{equation}\label{eq:eint}
\eps^B_\mathrm{int}[n;\xi_\mathrm{var}] = \frac{n}{2} \int_0^\infty g^A(r;\xi_\mathrm{var})\,V^B\big(r; \xi^B)\,2\pi r\,\mathrm{d}r,
\end{equation}
where $V^B(r; \xi^B)$ is the two-body interaction potential of ``B'' and $\xi^B$ is a dimensionless coupling constant of ``B'' (e.g. for 2DDFG, $\xi^B$ can be taken as $g_d \equiv m D^2 k_F/\hbar^2$). Minimizing $\eps^B[n;\xi_\mathrm{var}] \equiv \eps^B_\mathrm{K}[n;\xi_\mathrm{var}] + \eps^B_\mathrm{int}[n;\xi_\mathrm{var}]$ with respect to $\xi_\mathrm{var}$ at fixed density, we obtain (a) a variational upper bound for the ground state energy of ``B'', and (b) a mapping $\phi: \xi^B \rightarrow \xi^A$ that associates the ground states of ``A'' with the (approximate) ground states of ``B''. We refer to this scheme as  variational mapping, and the results obtained using the wave functions of ``A'' as \VM{A}. Note that only the knowledge of $g^A(r;\xi^A)$ and $\eps^A_{\mathrm{K}}[n;\xi^A]$ is required. Furthermore, if the microscopic interaction is a power-law function, the kinetic energy density can be extracted from the total energy using the Virial theorem~\cite{Skinner2010,SM}.

The quantum HV freezing criterion allows us to the estimate the freezing point of the target system ``B'' once the variational mapping $\phi$ to the ground states of a reference system is constructed: if the freezing transition occurs at $\xi^A_c$ in the reference system, it (supposedly) fulfills the HV criterion. Since the same wave function is variationally associated to the target system at $\xi_c^B \equiv \phi^{-1}(\xi_c^A)$, it will also satisfy the HV criterion at $\xi_c^B$.

As a final remark, we note that a necessary condition for the applicability of the variational mapping method is the convergence of the integral in Eq.~\eqref{eq:eint}. For large $r$, $g^A \approx 1$ and the integral converges provided that $V^B(r)$ falls faster than $1/r^2$. If $V^A(r) \sim 1/r^{n}$ with $n\geq 3$, the solution of the two-body Schr\"odinger's equation shows that $g^A(r)$ and its derivatives vanish at $r=0$ to all orders. Therefore, the small-$r$ convergence of the integral is guaranteed for all target potentials with a power-law repulsive core with finite $n$ (hence, excluding the hard-core gas). Finally, if $V^A(r) \sim 1/r$, $g^A \sim r^2$ for small $r$~\cite{Gori2004} and convergence requires the repulsive core of $V^B$ to softer than $1/r^4$.


\section{Variational mapping from 2DEG to 2DDFG}
\begin{figure}[t!]
\center
\includegraphics[scale=0.60]{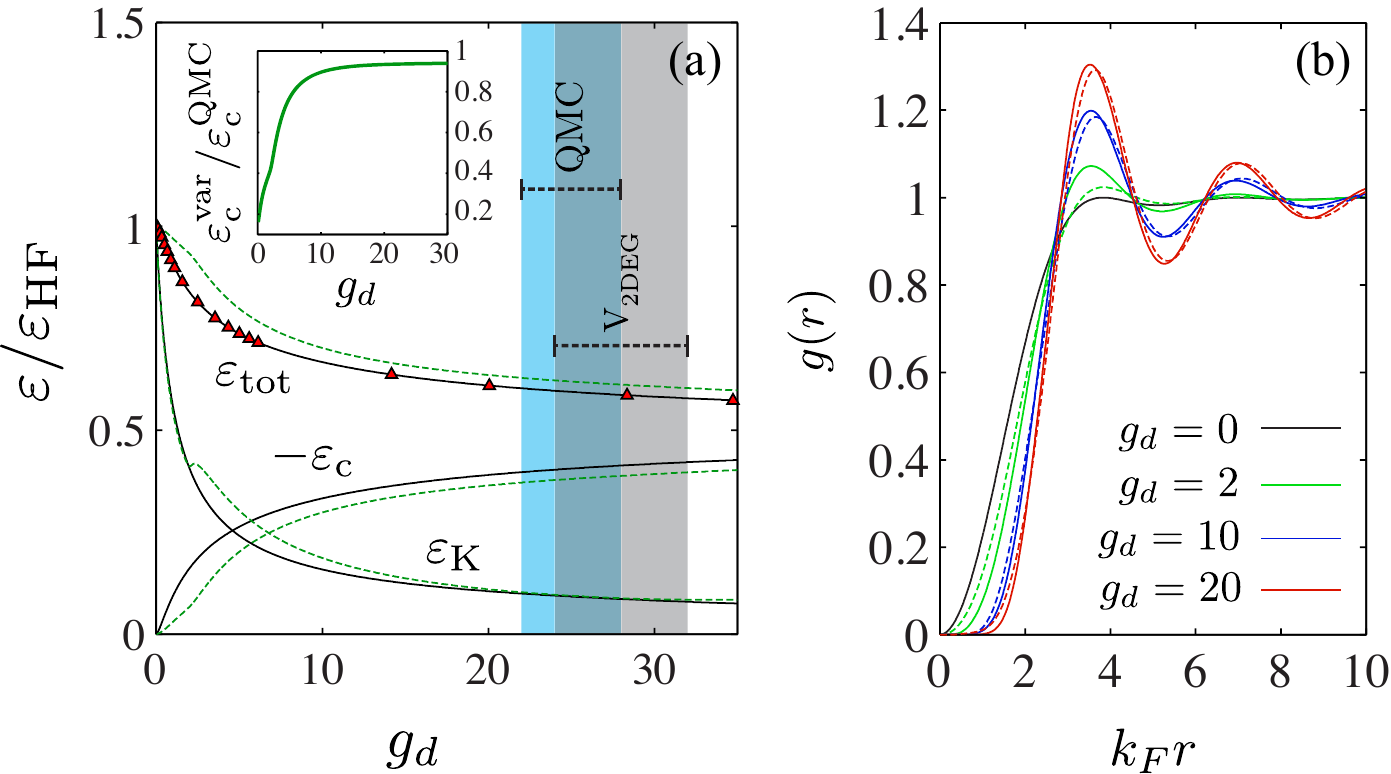}
\caption{(Color online) (a) The correlation ($\eps_\mathrm{c}$), kinetic ($\eps_\mathrm{kin}$), and total ($\eps_\mathrm{tot}$) energy per particle of 2DDFG in the units $\eps_\mathrm{HF} = (\hbar^2 \pi n/m)[1 + 128 g_d/(45 \pi)]$ \cite{SM,Matveeva2012}. The red triangles show the QMC results of Ref.~\cite{Matveeva2012}, the black solid lines are analytic fits. The green dashed lines are variational results based on 2DEG wave functions (\VM{2DEG}). The cyan bar shows the WC transition region as predicted by QMC, $g_d = 25 \pm 3$. The gray vertical bar is the variational estimate based on \VM{2DEG} and the Hansen-Verlet criterion, $g_d = 29 \pm 4$. The inset plot shows the ratio of the correlation energy of \VM{2DEG} over QMC. (b) The PDF $g(r)$ of 2DDFG from QMC (solid lines) and \VM{2DEG} (dashed lines). The black, green, blue and red lines correspond to $g_d = 0~(\text{Hartree-Fock}), 2, 10, 20$.} 
\label{fig:2DEG}
\end{figure}

The dipolar gas is assumed, for the moment, to be single-component and with an interaction law $V(r) = D^2/r^3$. The strength of dipolar interactions can be parametrized using the dimensionless coupling constant $g_d \equiv m D^2 k_F / \hbar^2$. We use ferromagnetic 2DEG ground-state energies from Ref.~\cite{Attaccalite2002} and the analytical representation of the 2DEG PDF given in Ref.~\cite{Gori2004}. Fig.~\ref{fig:2DEG}a shows the variationally obtained energies (green dashed lines) along with the QMC result from Ref.~\cite{Matveeva2012} (solid lines). The inset plots shows the fraction of the captured correlation energy. We find that the 2DEG wave functions remarkably represent more than $95\%$ of the correlation energy of 2DDFG in the strongly correlated regime ($g_d > 20$). The difference between the exact 2DDFG PDFs and the variationally associated 2DEG PDFs (Fig.~\ref{fig:2DEG}b) is barely noticeable for large $g_d$. The plots of $\varepsilon_\mathrm{K}$ and $\varepsilon_\mathrm{int}$ (Fig. 2a) indicate that the main overestimation comes from the interaction energy part. The error, however, is reduced for larger $g_d$, which indicates that the small-$r$ mismatch of the dipolar and Coulomb potentials is masked more effectively by the correlation hole.

In 2DEG, the transition from the ferromagnetic liquid to the WC phase takes place at $r_s = 28 \pm 3$~\cite{Drummond2009}. The 2DEG wave functions in this interval variationally map to $g_d = 28 \pm 4$ for 2DDFG, which we take as an estimate for the WC transition of 2DDFG using the HV criterion. This estimate is remarkably close to the QMC prediction $g_{d} = 25 \pm 3$~\cite{Matveeva2012}. The remarkable quality of the variationally obtained results in the strongly interacting regime and the decent estimate of the WC transition can be taken as a token of evidence for the presence of strong universal features in the strongly-correlated wave functions of the two models. Our estimate of the WC transition point is also a significant improvement over the available analytical estimates: the Hartree mean-field stability criterion predicts the transition at $g_d \approx 0.5$~\cite{Yamaguchi2010}, inclusion of exchange effects result in a slight improvement $g_d \approx 1.4$~\cite{Babadi2011b,Sieberer2011}, and inclusion of correlation effects using the Singwi-Tosi-Land-Sj\"olander scheme yields $g_d \approx 6$~\cite{Parish2012}.

\section{Quasi-2D dipolar Fermi gas} So far, we have only discussed the system of dipolar fermions in a strictly 2D configuration. This model can be experimentally simulated by optical or magnetic confinement of fermionic polar molecules~\cite{DipolarExp} or magnetic atoms~\cite{Burdick2012} about a plane, and polarizing the dipoles perpendicular to the confinement plane using an external dc field. Assuming a harmonic confining potential $U_\mathrm{trap} \approx m\omega_z ^2 z^2/2$, the 2D limit corresponds to the limit $\omega_z/\mu \rightarrow \infty$, where $\mu$ is the chemical potential. The trap frequency $\omega_z$ is finite in reality and the 2D layer has a finite width of the order of the transverse oscillator length $a_z \equiv [\hbar/(m\omega_z)]^{-1/2}$. Provided that $\hbar\omega_z > \mu$, only the lowest transverse band will be populated~\cite{Babadi2011a}. We refer to dipolar fermions in this setting as quasi-2D dipolar Fermi gas (q2DDFG). Finite transverse confinement modifies the short-range $r \lesssim a_z$ behavior of the effective two-body interactions, which is given by:
\begin{align}\label{eq:VQ2D}
\mathcal{V}^\mathrm{Q2D}_\mathrm{dip}(\rr) &= \int\mathrm{d}z\,\mathrm{d}z'\,|\phi_0(z)|^2\,|\phi_0(z')|^2\,\mathcal{V}_\mathrm{dip}^\mathrm{3D}(\rr,z-z'),
\end{align}
where $\mathcal{V}_\mathrm{dip}^\mathrm{3D}(\rr,z-z') = D^2\left(|\rr|^2 - 3z^2\right)/|\rr|^5$ is the dipole-dopole interaction in 3D space and $\phi_0(z) = e^{-z^2/(2 a_z^2)}/(\sqrt{\pi}\,a_z)^{\frac{1}{2}}$ is the transverse wave function of particles in the lowest band. The analytical expression for $\mathcal{V}^\mathrm{Q2D}_\mathrm{dip}(\rr)$ is given in the Ref.~\cite{Komenias2007} and yields:
\begin{equation}\label{eq:Vlimits}
\mathcal{V}^\mathrm{Q2D}_\mathrm{dip}(\rr) = \left\{
	\begin{array}{ll}
		[2D^2/(\sqrt{2\pi}a_z^3)]\ln(a_z/r) + \mathcal{O}(1) & r \lesssim a_z,\vspace{5pt}\\
		D^2/r^3 - 9 a_z^2/(2r^5) + \mathcal{O}(r^{-7}) & r \gtrsim a_z.
	\end{array}
\right.
\end{equation}
The much weaker repulsion in the region $r \lesssim a_z$ is due to the strong anisotropy of dipole-dipole interactions in 3D space. When the separation between the dipoles exceeds $a_z$, the ideal dipole-dipole interaction is asymptotically recovered. We use the variational mapping method to calculate the properties of q2DDFG using 2DDFG wave functions. In light of the analysis of the previous section, we expect the ground states wave functions of 2DDFG to comprise a decent set of variational trial states for q2DDFG since the long-range behavior of the two models asymptotically match.
\begin{figure}
\center
\includegraphics[scale=0.61]{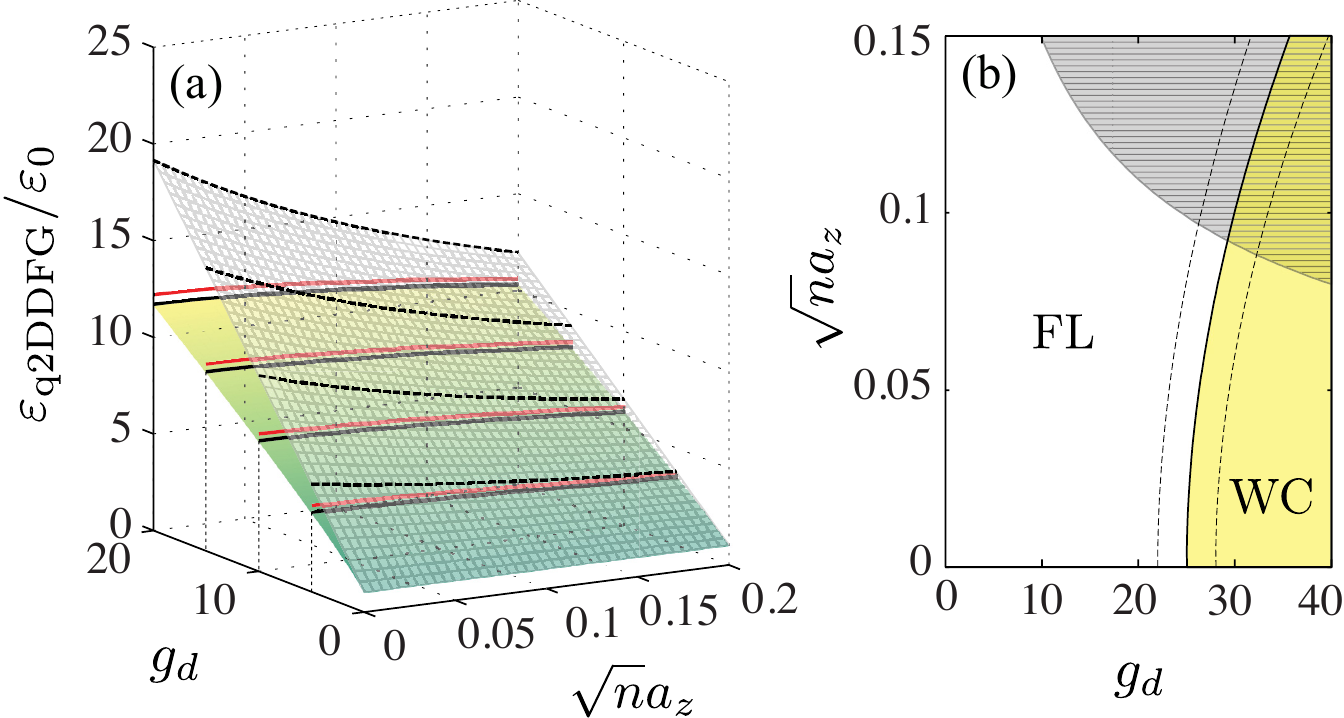}
\caption{(Color online) (a) The energy of quasi-2D dipolar Fermi gas (q2DDFG) as a function of dipole-dipole interaction strength $g_d$ and the layer width $a_z$ in the units of $\eps_0 \equiv \hbar^2 \pi n/m$. The upper surface and the accompanying black dashed lines show the result from Hartree-Fock theory, the lower surface and the black solid lines represent the variational result using 2DDFG ground states, and the solid red lines show the variational result using 2DEG ground states. The weak quadratic $a_z$-dependence of the variational results is a manifestation of the strong short-range correlations (b) The phase diagram of q2DDFG obtained using 2DDFG ground states and the quantum Hansen-Verlet freezing rule. FL and WC stand for Fermi liquid and Wigner crystal phases. The dashed lines indicate the lower and upper uncertainty bounds for the WC transition. The gray region in the top is where the single-band (quasi-2D) limit is not applicable anymore.}
\label{fig:DFG}
\end{figure}

The ground-states of q2DDFG are parametrized by two dimensionless quantities $(g_d,\tilde{a}_z)$, where $\tilde{a}_z = \sqrt{n}a_z$. Fig.~\ref{fig:DFG}a shows the variationally obtained energies along with the Hartree-Fock result. The phase diagram as a function of $g_d$ and $a_z$ is obtained using the HV criterion and is shown in Fig.~\ref{fig:DFG}b. We have also calculated the variational energies using 2DEG wave functions separately for comparison (shown as red lines in Fig.~\ref{fig:DFG}a). We find that both variational estimates lie remarkably close to each other and follow the same trend as a function of $g_d$ and $a_z$. The estimated energies based on 2DEG wave functions consistently lie slightly above those based on 2DDFG, as expected.

An interesting consequence of correlations is the qualitatively different dependence of the ground state energy on $a_z$, as compared to the Hartree-Fock theory. Indeed, for small $\sqrt{n}a_z$, Hartree-Fock theory predicts a linear dependence of the total energy on $a_z$ (see the black dashed lines in Fig.~\ref{fig:DFG}a; the exact Hartree-Fock energy expression is also given in Ref.~\cite{SM}), while both \VM{2DDFG} and \VM{2DEG} strongly suggest a quadratic $a_z$-dependence (see the solid black/red lines in Fig.~\ref{fig:DFG}a). In fact, \VM{2DDFG} energies can be parametrized to an excellent approximation as $\eps_\mathrm{q2DDFG}(g_d,a_z) \approx \eps_{\mathrm{2DDFG}}(g_d) - (\hbar^2 \pi n^2)\,(c_0 + c_1 g_d)\,a_z^2/m$ in the parameter regime $5<g_d<30$ and $\sqrt{n} a_z<0.1$, where $c_0 = 12.8$ and $c_1 = 1.45$ as obtained by fitting, and $\eps_{\mathrm{2DDFG}}(g_d)$ is given in Ref.~\cite{Matveeva2012} (cf. Ref.~\cite{SM} for an analytical fit). The quadratic dependence of $\eps_\mathrm{q2DDFG}$ on $a_z$ can be understood in simple terms: for large $g_d$, the short-range part of the effective dipole-dipole interaction is masked by the exchange-correlation hole and the $a_z$-dependence of the energy results from the leading correction to the quasi-2D interaction law in the large-$r$ limit, which is $\sim a_z^2/r^{5}$ (see Eq.~\ref{eq:Vlimits}). The WC transition line is parametrized as $g_d \approx 25 + c_2 n a_z^2,$, where $c_2 = 4.82 \times 10^2$. In contrast, the Hartree-Fock theory again spuriously predicts a linear $a_z$-dependence for the liquid stability phase boundary~\cite{Babadi2011b}.

\section{Experimental observation of correlation effects in q2DDFG}
\begin{figure}
\center
\includegraphics[scale=0.55]{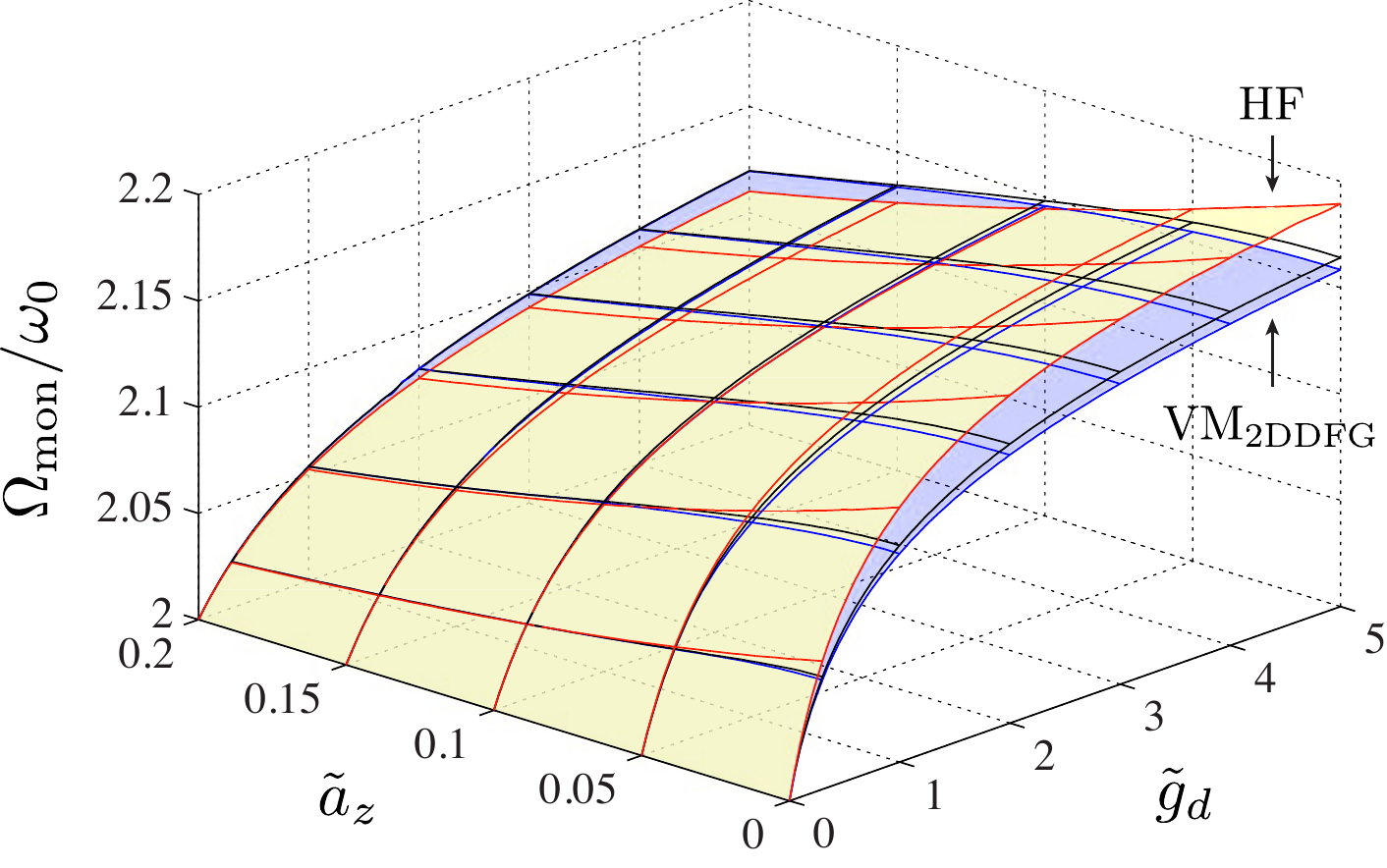}
\caption{(Color online) The monopole oscillation frequency $\Omega_\mathrm{mon}$ of q2DDFG in an isotropic trap at $T=0$. The yellow surface and red lines show the mean-field (HF) result. The blue surface and blue lines indicate the lower bound to $\Omega_\mathrm{mon}$ once correlations are included using the \VM{DFG} scheme. The upper bound is shown as black lines. The lower and upper bounds are obtained from ideal hydrodynamic theory and sum rules respectively. Inclusion of correlations changes the $a_z$-dependence of $\Omega_\mathrm{mon}$ significantly.}
\label{fig:mon}
\end{figure}

A prominent aspect of experiments with ultracold quantum gases is the possibility of carrying out direct and precise measurement of important quantities such as the static and dynamic structure factors~\cite{Veeravalli2008}, equation of state~\cite{Houcke2012} and the energy of collective modes in traps~\cite{Altemeyer2007}. Here, we show that the effects of strong correlations can be directly observed by measuring collective monopole (breathing) oscillation frequency $\Omega_\mathrm{mon}$ of q2DDFG in shallow traps.

We consider a q2DDFG in a harmonic in-plane trap potential $U_{xy} = m \omega_0^2\,(x^2+y^2)/2$, where $\omega_0 \ll \mu$. The collective excitations can be studied within the local density approximation (LDA) scheme. The equilibrium state of the trapped gas is obtained from balancing the trap restoring force and the pressure gradient, i.e. $\partial_\rr P[n_\mathrm{eq}(\rr)] + n_\mathrm{eq}(\rr)\,\partial_\rr U_\mathrm{trap}(\rr) = 0$. Here, $\neql(\rr)$ is the local equilibrium density and $P[n]$ is the LDA pressure functional. At zero temperature, the pressure is given by $P[n] = \int_0^{n} \mathrm{d}n'\,n'\,\partial(\mu[n'])/\partial_{n'}$, where $\mu[n] = \partial(n \eps)/\partial n$ is the chemical potential. The equilibrium condition yields $\mathrm{d}\neql/\mathrm{d}r = -m \omega_0^2 \kappa[\neql]\neql^2 r$, where $\kappa[n] \equiv n^{-2} [\partial^2(n\eps[n])/\partial^2 n]^{-1}$ is the compressibility, which can be calculated from the energies presented earlier. The density profile in the trap is obtained by solving the LDA equilibrium equation under the global particle number constraint, $\int\mathrm{d}^2\rr\,\neql(r) = N$. Correlations reduce the pressure at a given density, making the gas more compressible and resulting in a consistently smaller equilibrium radius of the trapped gas as compared to mean-field theory.

An exact treatment of collective oscillations at $T=0$ requires solving the Landau kinetic equation which is not feasible without the knowledge of the Landau parameters. Nevertheless, a lower bound to the frequency of collective oscillations can be found using the ideal hydrodynamic approximation. At zero temperature, the hydrodynamical description of the gas is provided by the conservation laws for the mass and momentum currents, i.e. $\partial_t n + \partial_\rr(n\mathbf{v})=0$ and $m(\partial_t\mathbf{v} + \mathbf{v}\cdot\partial_\rr\mathbf{v}) = -n^{-1}\partial_\rr P - \partial_\rr U_\mathrm{trap}$, where $\mathbf{v}$ is the macroscopic velocity field. Linearizing these equations about the equilibrium state and solving for $\delta n \equiv n - \neql$ gives
\begin{equation}\label{eq:hd}
\partial_t^2\,\delta n + \partial_\rr \cdot\left[\neql \partial_\rr\left(\frac{\delta n}{m\,\kappa[n_\mathrm{eq}]\,\neql^2}\right)\right]=0.
\end{equation}
The absence of dynamical Fermi surface deformations in the hydrodynamic approximation (which strictly increase the sound velocity in the case of isotropic repulsive interactions) implies that the obtained collective excitation energies are strictly lower than the exact values~\cite{HydroBound}. On the other hand, a rigorous upper bound is given by $\left(\hbar\Omega_\mathrm{mon}\right)^2 \leq m_\mathcal{M}^{(3)}/ m_\mathcal{M}^{(1)}$, where $m_\mathcal{M}^{(1)}$ and $m_\mathcal{M}^{(3)}$ are the first and third moments of the monopole response function and can be evaluated using exact sum rules~\cite{Pitaevski2003}. The numerical method for solving Eq.~(\ref{eq:hd}) and calculating the required moments are discussed in detail in Ref.~\cite{SM}. We calculate the mean-field $\Omega_\mathrm{mon}$ by solving the Boltzmann-Vlasov equation using the numerically exact method given in Ref.~\cite{Babadi2012}.

Fig.~\ref{fig:mon} shows the obtained results. We notice that the different $a_z$-dependence of the mean-field vs. correlated theory pointed out earlier also persists in $\Omega_\mathrm{mon}$. For small values of $\bar{a}_z$, correlations have a tendency to decrease the frequency of oscillations as compared to the prediction of the mean-field theory. This scenario is reversed, however, as $\bar{a}_z$ is increased. This reversal can be understood as a consequence of the weak dependence of the correlated theory on $a_z$ in contrast to the erroneously strong dependence of the mean-field theory. Although the results presented here correspond to the $T=0$ limit, the correlation effects are expected to persist as long as $T<T_F$. Suitably large dipolar interactions for observing such strong correlation effects are expected to be achievable in experiments with polar molecules such as KRb and NaK~\cite{DipolarExp}.

By modifying the shape of dipolar interactions using microwave fields, a wide gamut of soft repulsive potentials ranging from $1/r^6$ to $1/r^3$ is experimentally accessible~\cite{Astrakharchik2007}, allowing a direct experimental test of the universal behavior by directly measuring the structure factor using Bragg spectroscopy~\cite{Veeravalli2008}. The variational mapping method may be extended to superfluid and multi-component Fermi gases, which might be useful in investigating magnetism and superconductivity in electronic systems using dipolar gases as an experimental testbed.

\section{Acknowledgements} The authors would like to thank N. Matveeva and S. Giorgini for kindly providing us their QMC data. M. B and E. D. acknowledge the support from Harvard-MIT CUA, DARPA OLE program, AFOSR Quantum Simulation MURI, AFOSR MURI on Ultracold Molecules. This work was supported partially by the NSF through the University of Minnesota MRSEC under Award Number DRM-0819885.  M. F. is supported by UCOP.


%
%
%
%
%
%
%
%
%
%
%
%
%
%
%
%
%
%

\section{Supplemental Materials}

\subsection{1. The effective quasi-two-dimensional dipole-dipole interaction}
The formal expression for $\mathcal{V}^\mathrm{Q2D}_\mathrm{dip}$(r) was given in the main text in the integral form. The integration can be done analytically and we find $\mathcal{V}^\mathrm{2D}_\mathrm{dip}(r) = [D^2/(2\sqrt{2\pi}a_z^3)]v(r/a_z)$, where:
\begin{equation}\label{eq:u}
v(x) = e^{x^2/4}\left[(2 + x^2)\,K_0(x^2/4) - x^2 K_1(x^2/4)\right],
\end{equation}
and $K_{n}(x)$ denotes the modified Bessel function of the second kind of order $n$. For small $x$, we find:
\begin{equation}
v(x) = -4 - 2 \gamma - 2\ln(x^2/8) + \mathcal{O}(x^2),
\end{equation}
while for large $x$, we obtain the asymptotic expansion:
\begin{equation}
v(x) = 2\sqrt{2\pi}\left[\frac{1}{x^3} - \frac{9}{2x^5} + \mathcal{O}(x^{-7})\right].
\end{equation}
The Hartree-Fock energy density at $T=0$ can be easily calculated as:
\begin{align}\label{eq:ehf}
\eps^\mathrm{HF} &= \eps_0 + \frac{1}{4} \int_0^\infty x \left(1 - \frac{4 J_1(x)^2}{x^2}\right)\,\mathcal{V}^\mathrm{2D}_\mathrm{dip}\left(\frac{x}{\sqrt{4\pi n}}\right)\nonumber\\
&= \eps_0\left[1 + \frac{128}{45 \pi}\,g_d\,I(\bar{a}_z)\right],
\end{align}
where $\eps_0 = \hbar^2 \pi n/m$ in the kinetic energy of a non-interacting spin-polarized 2D Fermi gas, and $g_d$ and $\bar{a}_z$ were defined in the main text. $I(\bar{a}_z)$ can be expressed in terms of special functions:
\begin{multline}
I(\bar{a}_z) = \frac{5}{2}\,\,{}_2 \mathrm{F}_2\left(\frac{3}{2},2; \frac{5}{2},\frac{5}{2}; 8\pi\bar{a}_z^2\right)\\
+ \frac{45}{8192\,\pi^2\,\bar{a}_z^5}\Bigg[8\sqrt{2}\pi\bar{a}_z^2\,\Big(\mathrm{Ei}(8\pi\bar{a}_z^2)- \log(8\pi\bar{a}_z^2)\Big)\\
- \sqrt{2}\,(16\pi\bar{a}_z^2-3)\,e^{8\pi\bar{a}_z^2}\,\mathrm{Erf}\Big(\sqrt{8\pi}\bar{a}_z\Big)+ 32\sqrt{2}\,\pi^2\bar{a}_z^4\\
- 8\sqrt{2}\,(\gamma-3)\,\pi\bar{a}_z^2 - 24\,\bar{a}_z - 3\sqrt{2}\,\big(e^{8\pi\bar{a}_z^2}-1\big)\Bigg],
\end{multline}
where $\mathrm{Ei}$ is the exponential integral and ${}_2\mathrm{F}_2$ is the generalized hypergeometric function. Expanding $I(\bar{a}_z)$ for small $\bar{a}_z$, we get:
\begin{equation}
\eps^\mathrm{HF}/\eps_0 = 1 + \frac{128 g_d}{45\pi} - 2\sqrt{2} g_d\,\bar{a}_z + \mathcal{O}(g_d\bar{a}_z^2).
\end{equation}
Note the linear dependence of the Hartree-Fock energy on $a_z$.
 
\subsection{2. Remarks on the application of the variational mapping method to 2DDFG}
The application of the variational mapping method to 2DDFG is similar to mapping to 2DEG and closely resemble the treatment given in Ref.~\cite{Skinner2010SM}. As mentioned in the main text, this requires the knowledge of (1) the kinetic energy density, and (2) the pair-distribution function of the reference system. As a first step, we parametrize the FN-DMC energies of 2DDFG reported in Ref.~\cite{Matveeva2012SM}. Motivated by the weak-coupling perturbative analysis of 2DDFG given in Ref.~\cite{Lu2012SM}, and the known functional form of the energy density on $g_d$ in the crystal phase~\cite{Matveeva2012SM}, we parametrize the 2DDFG correlation energy using the following ansatz: 
\begin{multline}\label{eq:fitec}
\eps_c(g_d)/\epsilon_0 = (A g_d^2 + B g_d^3 + C g_d^4)\\
\times \log\left[1 + \frac{1}{c_1 g_d + c_{3/2} g_d^{3/2} + c_2 g_d^2 + c_3 g_d^3}\right].
\end{multline}
In the limit $g_d \ll 1$, the above ansatz yields $\eps_c \approx -A g_d^2 \log(c_1 g_d) + \mathcal{O}(g_d^{5/2})$, which is in agreement with the perturbation analysis of Lu and Shlyapnikov~\cite{Lu2012SM}, who find $A = -1/4$ and $c_1 \approx 1.43$. Here, we do not fix these parameters and keep both as fitting parameters.  Although we are only concerned with the liquid phase, the above ansatz also assumes the right expected form $\sim \alpha/\sqrt{g_d} + \beta + \gamma\,g_d$ in the crystal phase~\cite{Matveeva2012SM}. Optimizing the fit parameters in the range $g_d \in [0, 28]$, we find $A = -0.2511$, $B = -0.192$, $C=-0.01093$, $c_1 = 1.492$, $c_{3/2} = 0.004919$, $c_2 = 0.5814$ and $c_3 = 0.02382$. Note that the optimized values of $A$ and $c_1$ are very close to the known analytical values. The fit quality is excellent, with a root mean square error of $5 \times 10^{-4}$.

We extract the kinetic energy density of 2DDFG from its total energy density using the Virial theorem as follows: the general Virial theorem due to Cottrell and Paterson~\cite{Cottrell1951SM} implies $2\eps_\mathrm{K} - \frac{1}{N}\langle\sum_{i=1}^N \rr_i\cdot\partial \hat{\mathcal{V}}/\partial \rr_i\rangle + \ell\,\partial\eps/\partial\ell=0,$
where $\hat{\mathcal{V}}$ is the two-body interaction operator, $N$ is the number of particles and $\ell$ is a linear dimension of the system size. Strictly in 2D, dipole-dipole interaction is a homogeneous function of order $-3$ and the Virial term evaluates to $3\,\eps_\mathrm{int}$. Noting that $g_d \propto \ell^{-1}$, we get $2\eps_\mathrm{K} + 3\eps_\mathrm{int} = 2\eps + g_d \partial_{g_d}\eps[g_d]$. Using $\eps = \eps_\mathrm{K} + \eps_\mathrm{int}$, we finally get:
\begin{equation}
\eps_\mathrm{K} = \eps - g_d\partial_{g_d}\eps(g_d).
\end{equation}

The pair-distribution function of 2DDFG is reported in Ref.~\cite{Matveeva2012SM} for several values of $g_d$, using which we constructed an analytical representation for it in the spirit of the work by Gori-Giorgi {\it et al.}~\cite{Gori2004SM}. We will report the technical details of this procedure in a separate work.

\subsection{3. q2DDFG in isotropic traps}
The equilibrium density of the trapped gas can be obtained by solving Eq.~(6) in the main text. For a non-interacting gas, $\eps[n] = \eps_0 = \pi\hbar^2 n/m$ and we find:
\begin{equation}
n_{\mathrm{eq},0}(r) = \frac{m^2 \omega_0^2}{4\pi\hbar^2}\,(\rhom^2 - r^2),
\end{equation}
where $\rhom = (2N)^\frac{1}{4} \sqrt{2}\,[\hbar/(m\omega_0)]^{1/2}$ is the Thomas-Fermi radius of the cloud and $N$ is the number of particles. In the presence of interactions, the differential equation for $n_\mathrm{eq}$ may only be solved numerically. We arrive at a solution for $n_\mathrm{eq}$ by first finding upper and lower bounds to the density at center of the trap, $n_0$ using a direct search, and then solving for the function $n_\mathrm{eq}(r)$ that satisfies the global particle number constraint, $\int\mathrm{d}^2\rr\,\neql(r) = N$, by bisecting the bounding interval. Convergence within a relative error tolerance of $10^{-6}$ is achieved usually within 20 bisections.

\begin{figure}
\center
\includegraphics[scale=0.72]{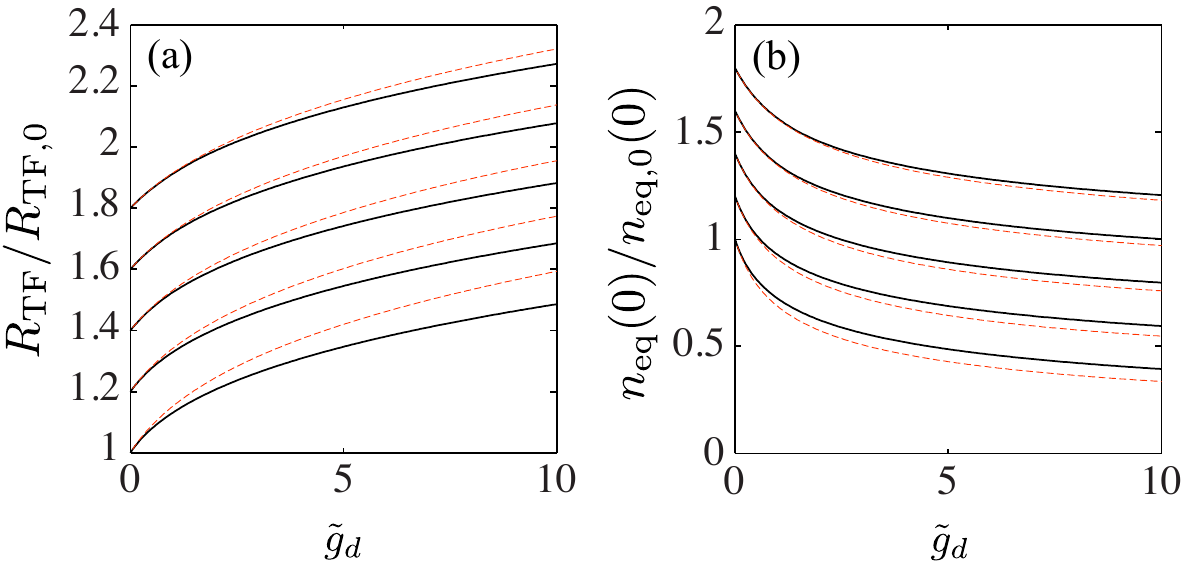}
\caption{The equilibrium radius (a) and density at the center of the trap (b) of q2DDFG at $T=0$ as a function of $\tilde{g}_{d}$ for various quasi-two-dimensionality factors (from bottom to top, $\sqrt{2\pi}\tilde{a}_z = 0, 0.1, 0.2, 0.3, 0.4$). The plots are shifted up in $0.2$ increments for better visibility. The solid and dashed lines correspond to the correlated and mean-field results respectively. Note that $R_{\mathrm{TF},0} = (8N)^{1/4} [\hbar/(m\omega_0)]^\frac{1}{2}$ and $n_{\mathrm{eq},0}(0) = m\omega_0 (2N)^{1/2}/(2\pi \hbar^2)$.}
\label{fig:eq}
\end{figure}

Fig.~\ref{fig:eq} shows (a) the equilibrium radius of the cloud (b) the density at the center of the trap. The plots shows both the results based on Hartree-Fock LDA energy density functional [Eq.~\ref{eq:ehf}] (red dashed lines) and the correlated energy obtained by variationally mapping to 2DDFG (black lines). Correlations make the gas more compressible and results in (1) larger density of particles at the center of the trap, and (2) smaller equilibrium radius.\\

The linearized hydrodynamic equation for the density variations (Eq.~7 in the main text) can be solved as follows. First, we note that the equation can be put in a self-adjoint form by changing variables to $\Psi \equiv \big[m\kappa_\mathrm{eq} \neql^2\big]^{-1/2}\delta n$~\cite{Csordas2006SM}. We then obtain $(\partial_t^2 + \hat{\GG})\Psi = 0$, where $\hat{\GG} = -\Beq\partial_\rr\cdot\left\{\neql\partial_\rr[\Beq\Psi]\right\}$ and $\Beq = (m\kappa_\mathrm{eq} \neql^2)^{-1/2}$. It is easily verified that $\hat{\GG}$ is self-adjoint with respect to the inner-product $\langle \Psi_1 | \Psi_2 \rangle = \int_{\Omega_+}\mathrm{d}^2\rr \,\Psi_1^*(\rr)\Psi_2(\rr)$, where $\Omega_+ = \{\rr : \neql(\rr) > 0\}$. Moreover, $\hat{\GG}$ is a Hermitian operator on the $L^2$ functions that remain finite on the boundary of $\Omega_+$.

In the absence of interactions, we can find the eigenfunctions of $\hat{\GG}$ analytically. Here, we are interested in monopole oscillations and hence, only the zero angular momentum solutions are needed. The (unnormalized) eigenfunctions are given by:
\begin{equation}
\Psi_n^{(0)}(r) = {}_2 F_1(-n, n+1, 1, r^2/R_{\mathrm{TF},0}^2),
\end{equation}
corresponding to the eigenvalue $\nu_n = \omega_0 \sqrt{2n(n+1)}$, where $n \in \mathbb{N}$. In the presence of interactions, the eigenfunctions can no longer be found analytically. We find the spectrum of $\hat{\GG}$ by calculating the matrix elements of $\hat{\GG}$ in the non-interacting basis and diagonalizing it. The spectrum consists of a single unphysical solution, $\Psi \sim \mathcal{B}_\mathrm{eq}^{-1}$ with eigenvalue 0. The oscillation frequency of the lowest lying physical monopole mode, $\Omega_\mathrm{mon}$, corresponds to the smallest non-zero eigenvalue. We found that truncating the basis set to the first 10 non-interacting eigenfunctions yields $\Omega_\mathrm{mon}$ to a relative accuracy level of $10^{-6}$.\\

We obtained the upper bound for $\Omega_\mathrm{mon}$ using the exact relation between the moments of the response function, i.e. $\left(\hbar\Omega_\mathrm{mon}\right)^2 \leq m_\mathcal{M}^{(3)}/ m_\mathcal{M}^{(1)}$. The monopole excitation operator and its corresponding response functions are $\hat{\mathcal{M}} = \sum_{i=1}^N r_i^2$ and $\chi_{\mathcal{M}}(\omega) = \hbar^{-1}\int\mathrm{d}t\,e^{i\omega t}\langle[\hat{\mathcal{M}}(t),\hat{\mathcal{M}}(0)\rangle_\mathrm{eq}$ respectively. The first and third moments can be calculated using sum rules~\cite{Pitaevski2003SM}:
\begin{align}
m_\mathcal{M}^{(1)} &= \langle 0|[[\hat{\mathcal{M}},\hat{\mathcal{H}}],\hat{\mathcal{M}}]|0\rangle/2,\nonumber\\
m_\mathcal{M}^{(3)} &= \langle 0|[[[\hat{\mathcal{M}},\hat{\mathcal{H}}],\hat{\mathcal{H}}],[\hat{\mathcal{H}},\hat{\mathcal{M}}]|0\rangle/2,
\end{align}
where $|0\rangle$ is the equilibrium state of the trapped gas. A direct calculation using commutation relations yield:
\begin{equation}
\frac{m_\mathcal{M}^{(3)}}{m_\mathcal{M}^{(1)}} = 2\omega_0\left[\frac{4(\mathcal{E}_\mathrm{K}+\mathcal{E}_\mathrm{trap}) + \mathcal{E}_{\mathcal{V}_2}}{8\,\mathcal{E}_\mathrm{trap}}\right]^\frac{1}{2},
\end{equation} 
where:
\begin{align}
\mathcal{E}_\mathrm{K} &= \int\mathrm{d}^2\rr\,\neql(\rr)\,\eps_\mathrm{K}[\neql(\rr)],\nonumber\\
\mathcal{E}_\mathrm{trap} &= \int\mathrm{d}^2\rr\,\neql(\rr)\,U_\mathrm{trap}(\rr),\nonumber\\
\mathcal{E}_{\mathcal{V}_2} &= \int\mathrm{d}^2\rr\,\neql(\rr)\,\eps_{\mathcal{V}_2}[\neql(\rr)].
\end{align}
Here, $\eps_{\mathcal{V}_2}[n]$ is given by:
\begin{multline}\label{eq:epsv2}
\eps_{\mathcal{V}_2}[n] = \frac{\hbar^2\pi n}{m}\frac{1}{\sqrt{8\pi}}\frac{g_d[n]}{(4\pi)^\frac{3}{2}\bar{a}_z[n]^3}\\
\times\int x\,{g}\big(x;g_d[n],\bar{a}_z[n]\big)\,v^{(2)}\Big(\frac{x}{\sqrt{4\pi}\bar{a}_z[n]}\Big).
\end{multline}
In the above equation, $v^{(2)}(x) = x\,\partial_x v(x) + x^2\,\partial^2_x v(x)$, and $g_d[n]$ and $\bar{a}_z[n]$ are the local dipole-dipole interaction strength and quasi-two-dimensionality. In Eq.~(\ref{eq:epsv2}), the pair-distribution function is that of the 2DDFG that variationally maps to $(g_d[n],\bar{a}_z[n])$.

\end{document}